\documentclass{article}
\usepackage{graphicx}
\begin{document}
\title{\bf{W boson mass anomaly and noncontractibility of the physical space}}
\author{\bf{Davor Palle} \\
ul. Ljudevita Gaja 35, 10000 Zagreb, Croatia \\
email: davor.palle@gmail.com}
\maketitle
\begin{abstract}
{The CDF II detector at the Tevatron collider reported significant 
tension between the measurement of the W boson mass and the Standard Model
prediction, assuming that 125 GeV scalar discovered at the LHC is 
the Higgs boson. We calculate one loop corrections to the W boson mass
within the theory of noncontractible space without the Higgs boson.
It turns out that our theory provides better agreement with the
CDF II detector result than the Standard Model.}
\end{abstract}

\section{Introduction and motivation}
Despite the great success of the Standard Model($SM$) of particle 
physics, we are witnessing the growth of theoretical 
and phenomenological problems of the model.
The massless Dirac neutrinos, the absence of the dark matter particle
and the conserved baryon number are notorious cosmological difficulties
of the model. Tevatron top quark charge asymmetry anomaly and anomalies 
observed in the $B$ meson semileptonic decays appear to be a great challenge
for the $SM$.

The latest problem has emerged from the CDF II measurement of the 
W boson mass that is $7\sigma$ away from the SM prediction, assuming that
the 125 GeV scalar is the $SM$ Higgs scalar \cite{CDFII}.

The $SM$ electroweak interactions contain a large number of free parameters
from the Higgs potential and Yukawa couplings. This is the reason why
the Higgs mechanism breaks the gauge symmetry but does not solve 
the problem of the gauge boson and fermion masses that are all free 
parameters.

We proposed some time ago \cite{Palle1} a new symmetry breaking mechanism
under the hypothesis of noncontractibility of the physical space.
The theory is UV nonsingular and free of the $SU(2)$ global anomaly
(we introduced the abbreviation $BY$ for this theory \cite{Palle1}).
As a conseqence it contains light and heavy Majorana neutrinos.
All Dirac fermion mixing angles must fulfill the following relation:

\begin{eqnarray}
\Theta_{W}=2 (\Theta_{12}^{D}+\Theta_{23}^{D}+\Theta_{31}^{D}).
\end{eqnarray}

The mixing matrix of the light Majorana neutrinos is defined by its 
corresponding Dirac neutrino submatrix \cite{Palle1}.
However, in the case of the inverted
mass hierarchy, for example if $m_{\nu,1}^{M} > m_{\nu,2}^{M}$ for 
light Majorana neutrinos,
$m_{N,1}^{M} < m_{N,2}^{M}$ for heavy Majorana neutrinos and
$m_{\nu,1}^{D} < m_{\nu,2}^{D}$ for Dirac neutrinos, then the see-saw mechanism and the Euler
matrix imply $\Theta_{12}^{D}=-\Theta_{12}^{M}$:

\begin{eqnarray*}
\left( \begin{array}{cc}
\cos \Theta_{12} & \sin \Theta_{12} \\
-\sin \Theta_{12}  & \cos \Theta_{12} \end{array}
\right)
\left( \begin{array}{c}
u_{1} \\
u_{2} \end{array}
\right)
\Leftrightarrow
\left( \begin{array}{cc}
\cos \Theta_{12} & -\sin \Theta_{12} \\
\sin \Theta_{12}  & \cos \Theta_{12} \end{array}
\right)
\left( \begin{array}{c}
u_{2} \\
u_{1} \end{array}
\right).
\end{eqnarray*}

The present knowledge
of the neutrino mixing matrix \cite{PDG} and the above $SU(2)$ cancellation
condition favour the inverted mass hierarchy. It is interesting to note 
that the most recent NOvA results \cite{NOvA} also favour the inverted hierarchy.

The heavy Majorana neutrinos are candidate particles for cold dark matter.
The baryon number is violated as a consequence of the violation of the 
lepton number (Majorana neutrinos) and the preservation of the $B-L$ \cite{Kolb}.

The UV cut-off, defined by Wick's theorem \cite{Palle1}, is a measure
of noncontractibility of the physical space:

\begin{eqnarray}
\Lambda = \frac{\pi}{\sqrt{6}}\frac{2}{g} M_{W},\   
 e = g \sin \Theta_{W},\ \cos \Theta_{W} = \frac{M_{W}}{M_{Z}}.
\end{eqnarray}

Even the qualitative discussion of Dyson-Schwinger equations contributes to
our understanding of the fermion spectrum patterns \cite{Palle1}.

In this paper we test perturbatively our $BY$ theory in the situation where the
consequences and differences between symmetry breaking mechanisms of the $SM$ and 
the $BY$ theory are insurmountable.

\section{Muon lifetime and the W boson mass}
It was noticed long ago that very accurate measurement of the muon lifetime
and the perturbative calculations to high orders in the leptonic 
environment allow for a stringent test of the electroweak theories
\cite{Sirlin}.

The CDF II Collab. \cite{CDFII} used the parametrization of the W boson mass
\cite{Awramik} based on the most accurate radiative corrections to the
muon lifetime contained in $\Delta r$. $M_{W}$ is found as a solution
of the nonlinear algebraic equation: 

\begin{eqnarray}
M_{W}^{2}(1-\frac{M_{W}^{2}}{M_{Z}^{2}})=\frac{\pi\alpha}{\sqrt{2}G_{\mu}}
(1+\Delta r(M_{W},...)).
\end{eqnarray}

We shall compare $\Delta r^{(\alpha)}$ corrections evaluated in the $SM$ and
the $BY$ theory. On higher order corrections we shall comment later.
The overview of the one loop corrections the reader can find in \cite{Hollik}.
From our earlier loop computations in QCD \cite{Palle2} or in the
electroweak part \cite{Palle3} of the $BY$ theory, considerable deviations
from the $SM$ are found only if very heavy particles are present in the loops
(t-quark for example) or for large external momenta. 

The Higgs boson enters the loops of the weak boson self energies in the $SM$
while it is absent in the $BY$ theory. The scalar part of the $SM$ Lagrangian is:

\begin{eqnarray}
(D_{\mu}\Phi)^{\dagger}D^{\mu}\Phi-V(\Phi),\ 
V(\Phi)=\frac{\lambda}{4}(\Phi^{\dagger}\Phi)^{2}-\rho^{2}\Phi^{\dagger}\Phi,\ 
\Phi(x)=\left( \begin{array}{c}\phi^{+}(x) \\ \phi^{0}(x) \end{array}\right).
\end{eqnarray}

The symmetry is broken because of the peculiar potential $V(\Phi)$ resulting
in the nonvanishing vacuum expectation value $v=\frac{2\rho}{\sqrt{\lambda}}$ and
the scalar doublet looks as:

\begin{eqnarray}
\Phi(x)=\left( \begin{array}{c}\phi^{+}(x) \\ \frac{1}{\sqrt{2}}(v+H(x)+\imath \chi (x))
\end{array}\right).
\end{eqnarray}
 
The mass of the Higgs boson is $M_{H}=\sqrt{2}\rho$, whilst Nambu-Goldstone
scalars $\chi,\ \phi^{+},\ \phi^{-}$ have gauge dependent masses.

The $BY$ theory contains a scalar doublet with the following relations:

\begin{eqnarray}
\langle V(\Phi) \rangle =0 \Longleftrightarrow \langle \Phi^{\dagger}\Phi
\rangle=0.
\end{eqnarray}

The scalar doublet is now:

\begin{eqnarray}
\Phi(x)=\left( \begin{array}{c}\phi^{+}(x) \\ \frac{1}{\sqrt{2}}(v+\zeta(x)+\imath \chi (x))
\end{array}\right).
\end{eqnarray}

Since the Nambu-Goldstone particles are zero-norm states \cite{Kugo}
$\langle\phi^{+} |\phi^{+}\rangle=\langle\phi^{-}|\phi^{-}\rangle= 
\langle\chi|\chi\rangle=0$, from Eq.(6) we conclude that $\langle\zeta|\zeta\rangle=0$.
Thus, the particle $\zeta$ is a zero-norm state (unphysical particle)
with a vanishing mass $M_{\zeta}=0$.
The parameter $v=\frac{\sqrt{6}}{\pi}\Lambda$ is extracted from Eq.(2), 
i.e., the hypothesis of the noncontractible space \cite{Palle1}. The auxiliary
particle  $\zeta$ has obviously the same couplings to weak bosons 
like the Higgs boson in the $SM$. It is very well known that 
the Nambu-Goldstone particles ensure the gauge invariance of the 
amplitudes, while the $\zeta$ particle ensures the unitarity bounds on
the cross sections and the invariance on the regularization procedures in
the loop calculations.

Now we can procede with the contribution of the self energy (SE) diagrams
to the $\Delta r^{(\alpha)}$ \cite{Sirlin,Hollik}:

\begin{eqnarray}
\Delta r^{(\alpha)}_{SE}&=&\frac{\partial\Sigma^{\gamma}(q^{2})}{\partial q^{2}}(0)
+2\frac{c_{w}}{s_{w}}\frac{\Sigma^{\gamma Z}(0)}{M_{Z}^{2}}
+\frac{c_{w}^{2}}{s_{w}^{2}}(\frac{\Sigma^{W}(M_{W}^{2})}{M_{W}^{2}}
-\frac{\Sigma^{Z}(M_{Z}^{2})}{M_{Z}^{2}})  \nonumber \\
&&+\frac{\Sigma^{W}(0)-\Sigma^{W}(M_{W}^{2})}{M_{W}^{2}},
\end{eqnarray}

where we denote $c_{w}=\frac{M_{W}}{M_{Z}},\ s_{w}=\sqrt{1-c_{w}^{2}}$.

The loop-suppressed electroweak or QCD processes studied in refs.
\cite{Palle2,Palle3} within the $BY$ theory have a well defined 
limes $\Lambda \rightarrow +\infty$.
On the other hand the tree level definition of W boson mass by the 
UV cut-off $\Lambda$ in Eq.(2) and the algebraic equation for the
W boson mass in Eq.(3) require the regularization of the one and
two point Green's functions \cite{Bogoliubov} to achieve the correct
UV limes of Eq.(8)(for definitions of Green's functions see Appendix A).

One can find the explicit forms of the electroweak boson self energies
in ref.\cite{Consoli}. In Appendix B we have written down W boson self 
energy in our notations, as an example.

It is instructive to emphasize and isolate the contribution of the Higgs
boson ($\zeta$ particle) to the W and Z boson self energies in the $SM$
($BY$ theory):

\begin{eqnarray}
\Sigma^{Z}(q^{2})_{H}=-\frac{\alpha}{4\pi c_{w}^{2}s_{w}^{2}}
[B_{22}(q^{2};M_{H},M_{Z})-M_{Z}^{2}B_{0}(q^{2};M_{H},M_{Z})
-\frac{1}{4}A(M_{H})], \nonumber \\
\Sigma^{W}(q^{2})_{H}=-\frac{\alpha}{4\pi s_{w}^{2}}
[B_{22}(q^{2};M_{H},M_{W})-M_{W}^{2}B_{0}(q^{2};M_{H},M_{W})
-\frac{1}{4}A(M_{H})].
\end{eqnarray}

One can check the differences of the Higgs boson contributions to the
$\Delta r^{(\alpha)}$ at the second column of TABLE I. in \cite{Awramik} between 
two different Higgs masses with the differences evaluated by Eqs.(8) and (9).
The agreement between our calculations and \cite{Awramik} is at the level of pro mille.

Any vertex or box contribution with only one weak boson in the loop \cite{Sirlin,Hollik}
cannot be significantly changed by the introduction of the UV cut-off $\Lambda$.
The vertex with $W$ and $Z$ bosons is proportional to $B_{0}(q^{2}=0;M_{W},M_{Z})$
and it does not differ between $\Lambda=+\infty$ and $\Lambda<+\infty$ (see Appendix A).

The box diagrams with two weak bosons (see for example Eq.(4.13) of ref. \cite{Hollik})
in $BY$ contribute by the amount:

\begin{eqnarray}
\Delta r^{(\alpha)}_{(WZ\ box)}(BY)&=&-\frac{\alpha}{4\pi}(1-\frac{5}{s_{w}^{2}}
+\frac{5}{2 s_{w}^{4}})\ln (c_{w}^{2})\times J(\Lambda)/J(+\infty), \\
J(\Lambda)&=&\int^{+\Lambda^{2}}_{0}d y \frac{1}{(y+M_{W}^{2})(y+M_{Z}^{2})}.
\nonumber
\end{eqnarray}

The one and two loop QCD contributions $\Delta r^{(\alpha\alpha_{s})}$
and $\Delta r^{(\alpha\alpha^{2}_{s})}$ (see TABLE I of \cite{Awramik})
are insensitive to the cut-off $\Lambda$
at the small scale of the decay \cite{Palle2}.
Very small corrections to the $SM$ could be expected only from the t-quark in the
two loop electroweak contributions $\Delta r^{(\alpha^{2})}_{ferm}$.

Acknowledging the presented arguments, we may procede with the numerical
evaluations of $\Delta r^{(\alpha)}$ and $M_{W}$ in the $BY$ theory in the next chapter.

\section{Results and discussion}
Our first step is to evaluate the $SM$ prediction of $M_{W}(SM)$ using Eq.(6) and
coefficients from Eq.(9) of ref.\cite{Awramik} and the following input
parameters \cite{PDG}:

\begin{eqnarray}
\alpha&=&1/137.035999180,\ M_{H}=125.25\ GeV,\ M_{Z}=91.1876\ GeV, \nonumber \\
m_{t}&=&172.76\ GeV,\ d \alpha=0,\ d \alpha_{s}=0 \nonumber \\
&\Rightarrow& M_{W}(SM)=80.35846\ GeV.
\end{eqnarray}

From our Eq.(3), $M_{W}(SM)$ and $G_{\mu}=1.1663788\times 10^{-5}\ GeV^{-2}$ we find
$\Delta r (SM)=380.202\times 10^{-4}$.

We have to modify $\Delta r (BY)$ to one loop order by adding self energy
$\Delta r^{(\alpha)}_{SE} (BY)$ and $WZ$box diagrams $\Delta r^{(\alpha)}_{WZbox} (BY)$
contributions (Eqs.(8) and (10)) to the $\Delta r (SM)$ and subtracting the corresponding SM
one loop terms:
\begin{eqnarray}
\Delta r (BY;M_{W})&=&\Delta r (SM;M_{W}(SM))+\Delta r^{(\alpha)}_{SE} (BY;M_{W})
-\Delta r^{(\alpha)}_{SE} (SM;M_{W}) \nonumber \\
&&+\Delta r^{(\alpha)}_{WZ box} (BY;M_{W})
-\Delta r^{(\alpha)}_{WZ box} (SM;M_{W}).
\end{eqnarray}

We use the regularized Green's functions $A^{\Lambda}_{reg}(m),\ B^{\Lambda}_{0,reg}
(q^{2};m_{1},m_{2})$ (see Appendix A) and the massless auxiliary field $\zeta$
in the estimate of the $\Delta r^{(\alpha)}_{SE}(BY)$.

The nonlinear algebraic equation Eq.(3) for $M_{W}$ applied to $\Delta r (BY;M_{W})$
gives us the wanted $M_{W}(BY)$. Note that we have to treat even the UV cut-off 
as a function of $M_{W}$ (Eq.(2)), although with a small impact on the result.

We present the solution of Eq.(3) for $M_{W}(BY)$ and the corresponding
radiative corrections ($m_{b}=4.18\ GeV$, the following CKM matrix parametrization is applied
\cite{PDG}: $\sin \Theta_{12}=0.22650,\ \sin \Theta_{23}=0.04053,\ 
\sin \Theta_{13}=0.00361,\ \delta=1.196\ rad$; note that the light quark
contributions to the photon self energy is usually estimated from the 
experiment by a dispersion relation): 

\begin{eqnarray}
&&M_{W}(BY)=80.47547\ GeV,\ \Lambda=320.568\ GeV,\ 
\Delta r(BY)=304.992\times 10^{-4}, \nonumber \\
&&-----------------------------   \nonumber \\
&&\Delta r(BY;SE,ferm)=334.401\times 10^{-4},\ \Delta r(SM;SE,ferm)=340.891\times 10^{-4}, \nonumber \\
&&\Delta r(BY;SE,bos)=-92.543\times 10^{-4},\ \Delta r(SM;SE,bos)=-26.684\times 10^{-4}, 
\nonumber \\
&&\Delta r(BY;WZbox)=39.968\times 10^{-4},\ \Delta r(SM;WZbox)=42.829\times 10^{-4}.
\end{eqnarray}

One can observe that the largest differences between the $BY$ theory and the $SM$
comes from the self energy boson sector owing to the absence of the Higgs scalar.
We can verify this fact by Eqs.(8) and (9), namely:

\begin{eqnarray}
&&\Delta r^{(\alpha)}_{BY}(SE,M_{\zeta}=0,\Lambda=+\infty)_{\zeta} 
-\Delta r^{(\alpha)}_{SM}(SE,M_{H}=125.25\ GeV,\Lambda=+\infty)_{H} \nonumber \\
&&=-66.473\times 10^{-4}, \nonumber \\
&&\Delta r^{(\alpha)}_{BY}(SE,M_{\zeta}=0,\Lambda=320.568\ GeV)_{\zeta} \nonumber \\ 
&&-\Delta r^{(\alpha)}_{SM}(SE,M_{H}=125.25\ GeV,\Lambda=320.568\ GeV)_{H}
=-66.606\times 10^{-4}. \nonumber
\end{eqnarray}

The difference between the fermion self energy parts appears to be predominantly due to a very
high t-quark mass, precisely like in the B meson semileptonic decay anomalies \cite{Palle3}.
This can be easily verified numerically by reducing the t-quark mass.

The linear perturbation of Eq.(3) from the $SM$ to the $BY$ quantities
yields:
\begin{eqnarray}
&&\delta M_{W}=M_{W}(BY)-M_{W}(SM), \nonumber \\
&&\delta(\Delta r)=\Delta r(BY,M_{W}(SM))
-\Delta r(SM,M_{W}(SM)), \nonumber \\
&&\delta M_{W}=\frac{\pi \alpha}{2\sqrt{2}G_{\mu}M_{W}(SM)}
(1-2\frac{M_{W}^{2}(SM)}{M_{Z}^{2}})^{-1}\delta(\Delta r) \\
\Longrightarrow &&\delta M_{W}=0.11640\ GeV,\ M_{W}(BY)=80.47486\ GeV.
\end{eqnarray}

Unlike the $SM$, the $BY$ prediction for the W boson mass is larger than the
experimental average value \cite{CDFII} $M_{W}($CDFII$)=80.4335\ GeV$ by $3.76\sigma$,
anticipating the same theoretical uncertainty as in the $SM$ \cite{CDFII}.

The reanalysis of the ATLAS Collab. \cite{ATLAS} of the W boson mass
measurement at $\sqrt{s}=7\ TeV$, $M_{W}=80360 \pm 16\ MeV$,
reveals the astonishing difference from the CDFII measurement \cite{CDFII} at 
$\sqrt{s}=1.96\ TeV$ of $4\sigma$. The hadron colliders measurements of
the electroweak observables depend on the parton distribution functions (PDFs).
However, the PDFs are derived from the singular QCD ($\Lambda = \infty$).
The nonsingular QCD ($\Lambda = 320.6\ GeV$) gives modified PDFs at very 
small scales (very large momenta) \cite{Palle2}. We expect consequently
the larger difference between the $M_{W}(BY)$ and $M_{W}(ATLAS)$ than
between the $M_{W}(BY)$ and $M_{W}(CDFII)$. Indeed, $M_{W}(BY)$ is $6.76\sigma$
away from $M_{W}(ATLAS)$ and $3.76\sigma$ away from $M_{W}(CDFII)$.
It is not unexpected that the LHCb measurement \cite{LHCb} of the W boson mass at
$\sqrt{s}=13\ TeV$ gives even smaller $M_{W}(LHCb)=80354\ MeV$.

If we interpolate $M_{W}(\sqrt{s})$ with the Lagrange polynomials at three
measured $\sqrt{s}=1.96,7,13\ TeV$, one gets $M_{W}(0.2\ TeV)=80.4739\ GeV$
(see Fig.1).
At the scale $\sqrt{s}=0.2\ TeV$ and below, the QCD(SM) and the QCD(BY)
do not distinguish considerably \cite{Palle2}. The extrapolated $M_{W}(0.2\ TeV)$
is almost a perfect match to $M_{W}(BY)$.
\begin{figure}[htb]
\centerline{
\includegraphics[width=12cm]{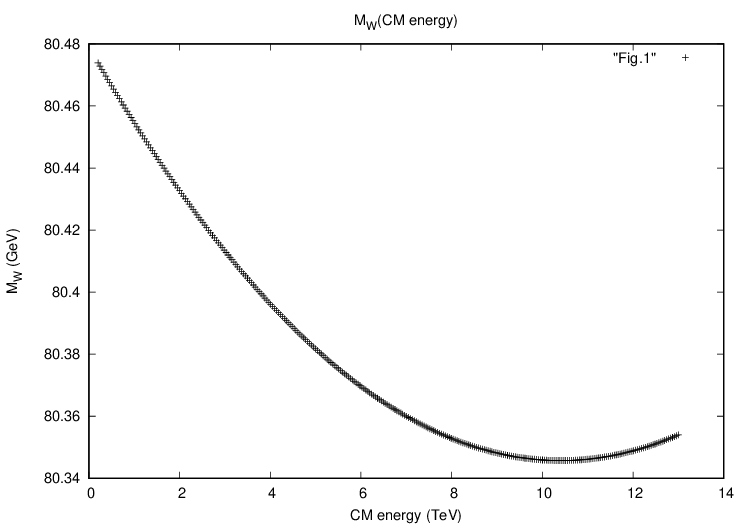}}
\caption{$M_{W}$(exp) as a function of $\sqrt{s}$ evaluated with the Lagrange polynomials
interpolation.}
\end{figure}

We would like to stress that the $BY$ theory cannot substantially improve the $SM$ 
calculation of the anomalous magnetic moment of the muon \cite{Palle4}
perturbatively. In addition, we point out that there is no lepton universality violation
in the $BY$ theory. The W boson scattering at the tree level in the SM deviates substantially
from the same process in the BY theory \cite{New1}.

Whether the $125\ GeV$ scalar is a Higgs scalar or just the scalar meson as 
a QCD bound state of mixed gluonium and toponium \cite{Cea,Palle5}, remains 
to be resolved by the experimental physics.

At the end, let us
emphasize the universality of our hypothesis of noncontractible
space that is testified within the Einstein-Cartan gravity and cosmology \cite{Palle6,New2,New3}.

\vspace{10mm}
{\bf Appendix A}
\vspace{5mm}
\newline
The standard definitions of one and two point
Green's functions are:

\begin{eqnarray*}
\frac{\imath}{16 \pi^{2}} A(m) =\mu^{4-D} \int \frac{d^{D}k}{(2 \pi)^{D}}\frac{1}{k^{2}-m^{2}}, 
\end{eqnarray*}
\begin{eqnarray*}
\frac{\imath}{16 \pi^{2}} B_{0;\mu;\mu\nu}(q^{2};m_{1},m_{2})
 =\mu^{4-D} \int \frac{d^{D}k}{(2 \pi)^{D}}\frac{1;k_{\mu};k_{\mu}k_{\nu}}{[k^{2}-m_{1}^{2}][(k+q)^{2}-m_{2}^{2}]}, 
\end{eqnarray*}
\begin{eqnarray*}
B_{\mu} = q_{\mu} B_{1},\ B_{\mu\nu} = g_{\mu\nu} B_{22}+q_{\mu}q_{\nu}B_{21},
\end{eqnarray*}
\begin{eqnarray*}
B_{1}(q^{2};m_{1},m_{2})=\frac{1}{2 q^{2}}[A(m_{1})-A(m_{2})
+(m_{2}^{2}-m_{1}^{2}-q^{2})B_{0}(q^{2};m_{1},m_{2})],
\end{eqnarray*}
\begin{eqnarray*}
B_{22}(q^{2};m_{1},m_{2})=&&\frac{1}{6}[A(m_{2})+2 m_{1}^{2}B_{0}(q^{2};m_{1},m_{2}) \\
&&+(q^{2}+m_{1}^{2}-m_{2}^{2})B_{1}(q^{2};m_{1},m_{2})]
+m_{1}^{2}+m_{2}^{2}-\frac{1}{3}q^{2}].
\end{eqnarray*}

We can calculate the integrals with the usual dimensional regularization
($D=4-\epsilon$):

\begin{eqnarray*}
A^{\infty}(m)&=&m^{2}(\Delta -\ln \frac{m^{2}}{\mu^{2}}+1),\ 
\Delta=\frac{2}{\epsilon}-\gamma+\ln(4\pi), \\
&&\Rightarrow A^{\infty}_{reg}(m)=m^{2}(-\ln \frac{m^{2}}{\mu^{2}}+1),
\end{eqnarray*}

\begin{eqnarray*}
&&B_{0}^{\infty}(p^{2};m_{1},m_{2})=\Delta+B_{0,reg}^{\infty}(p^{2};m_{1},m_{2}),\\
&&B_{0,reg}^{\infty}(p^{2};m_{1},m_{2})=
-\int^{1}_{0}d x \ln\frac{x^{2}q^{2}-x(q^{2}+m_{1}^{2}-m_{2}^{2})
+m_{1}^{2}}{\mu^{2}}.
\end{eqnarray*}

Analogous Green's functions for $\Lambda < +\infty$ appear in the well
know form \cite{Palle2,Palle3,Palle4}:

\begin{eqnarray*}
&&A^{\Lambda}(m)=-\Lambda^{2}+m^{2}\ln\frac{\Lambda^{2}+m^{2}}{m^{2}}
=\Delta_{\Lambda}+A^{\Lambda}_{reg}(m), \\
&&\Delta_{\Lambda}=-\Lambda^{2}+m^{2}\ln\frac{\Lambda^{2}+m^{2}}{\mu^{2}}
-m^{2}, \\
&& \Rightarrow A^{\Lambda}_{reg}(m)=A^{\infty}_{reg}(m),
\end{eqnarray*}
\begin{eqnarray*}
B_{0}^{\Lambda}(p^{2};m_{1},m_{2})=
\frac{1}{2}[\tilde{B}_{0}^{\Lambda}(p^{2};m_{1},m_{2})+
\tilde{B}_{0}^{\Lambda}(p^{2};m_{2},m_{1})],
\end{eqnarray*}
\begin{eqnarray*}
\tilde{B}_{0}^{\Lambda}(p^{2};m_{1},m_{2})=
[\int_{0}^{\Lambda^{2}}d y K(p^{2},y)+\theta (p^{2}-m_{2}^{2}) 
 \int_{-(\sqrt{p^{2}}-m_{2})^{2}}
^{0}d y \Delta K(p^{2},y) ]\frac{1}{y+m_{1}^{2}}, \\
K(p^{2},y)=\frac{2 y}{-p^{2}+y+m_{2}^{2}+
\sqrt{(-p^{2}+y+m_{2}^{2})^{2}+4 p^{2} y}},  \hspace*{40 mm}\\
\Delta K(p^{2},y)=\frac{\sqrt{(-p^{2}+y+m_{2}^{2})^{2}+4 p^{2} y}}{p^{2}}.
 \hspace*{60 mm}
\end{eqnarray*}

Acknowledging the relation:
\begin{eqnarray*}
\lim_{\Lambda \rightarrow \infty} [B_{0}^{\Lambda}(q^{2};m_{1},m_{2})
-B_{0}^{\Lambda}(0;m_{1},m_{2})]=B_{0,reg}^{\infty}(q^{2};m_{1},m_{2})
-B_{0,reg}^{\infty}(0;m_{1},m_{2}),
\end{eqnarray*}

we define the regularized function:
\begin{eqnarray*}
B_{0,reg}^{\Lambda}(q^{2};m_{1},m_{2})=B_{0}^{\Lambda}(q^{2};m_{1},m_{2})
-B_{0}^{\Lambda}(0;m_{1},m_{2})+B_{0,reg}^{\infty}(0;m_{1},m_{2}).
\end{eqnarray*}

As a consequence, it follows:
\begin{eqnarray*}
\lim_{\Lambda \rightarrow +\infty}B_{0,reg}^{\Lambda}(q^{2};m_{1},m_{2})
=B_{0,reg}^{\infty}(q^{2};m_{1},m_{2}),\ 
B_{0,reg}^{\Lambda}(0;m_{1},m_{2})=B_{0,reg}^{\infty}(0;m_{1},m_{2}).
\end{eqnarray*}

With equation
\begin{eqnarray*}
B_{1}(0;m_{1},m_{2})=-\frac{1}{2}B_{0}(0;m_{1},m_{2})
+\frac{1}{2}(m_{2}^{2}-m_{1}^{2})\frac{\partial B_{0}(q^{2};m_{1},m_{2})}
{\partial q^{2}}(0),
\end{eqnarray*}
we complete the list of all the Green's functions necessary for the
evaluation of the electroweak gauge boson self energies.

\vspace{10mm}
{\bf Appendix B}
\vspace{5mm}
\newline
Here we write down in our notation the
17 diagrams for W boson self energies depicted in ref.\cite{Consoli}
in the 't Hooft-Feynman gauge:

\begin{eqnarray*}
\Sigma^{W}(q^{2})&=&\sum_{j=1}^{17} \Sigma^{W}_{j}(q^{2}), \\
\Sigma^{W}_{1}(q^{2})&=&\frac{\alpha}{4\pi}\frac{1}{3s^{2}_{w}}
\sum_{l=1}^{3}[-\frac{1}{3}q^{2}-A(m_{l})-A(0)+m_{l}^{2}+
(q^{2}-\frac{1}{2}m_{l}^{2})B_{0}(q^{2};m_{l},0) \\
&&+\frac{m_{l}^{4}}{2 q^{2}}(B_{0}(0;m_{l},0)-B_{0}(q^{2};m_{l},0))] \\
&&+\frac{\alpha}{4\pi}\frac{1}{s^{2}_{w}}\sum_{i,k=1}^{3}|V_{ik}|^{2}
[-\frac{1}{3}q^{2}-A(m_{u,i})-A(m_{d,k})+m_{u,i}^{2}+m_{d,k}^{2} \\
&&+(q^{2}-\frac{1}{2}m_{u,i}^{2}-\frac{1}{2}m_{d,k}^{2})B_{0}(q^{2};m_{u,i},m_{d,k}) \\
&&+\frac{(m_{u,i}^{2}-m_{d,k}^{2})^{2}}{2 q^{2}}
(B_{0}(0;m_{u,i},m_{d,k})-B_{0}(q^{2};m_{u,i},m_{d,k}))], \\
\Sigma^{W}_{2}(q^{2})&=&-\frac{\alpha}{4\pi}\frac{1}{s^{2}_{w}}
B_{22}(q^{2};M_{Z},M_{W}),\ 
\Sigma^{W}_{3}(q^{2})=-\frac{\alpha}{4\pi}\frac{1}{s^{2}_{w}}
B_{22}(q^{2};M_{W},M_{H}), \\
\Sigma^{W}_{4}(q^{2})&=&\frac{\alpha}{4\pi}\frac{M_{W}^{2}}{s^{2}_{w}}
B_{0}(q^{2};M_{W},M_{H}),\
\Sigma^{W}_{5}(q^{2})=\frac{\alpha}{4\pi}M_{W}^{2}
B_{0}(q^{2};0,M_{W}), \\
\Sigma^{W}_{6}(q^{2})&=&\frac{\alpha}{4\pi}M_{W}^{2}\frac{s^{2}_{w}}{c^{2}_{w}}
B_{0}(q^{2};M_{Z},M_{W}),\
\Sigma^{W}_{7}(q^{2})=\frac{\alpha}{4\pi}\frac{1}{4 s^{2}_{w}}A(M_{H}), \\
\Sigma^{W}_{8}(q^{2})&=&\frac{\alpha}{4\pi}\frac{1}{4 s^{2}_{w}}A(M_{Z}),\ 
\Sigma^{W}_{9}(q^{2})=\frac{\alpha}{4\pi}\frac{1}{2 s^{2}_{w}}A(M_{W}), \\
\Sigma^{W}_{10}(q^{2})&=&\frac{\alpha}{4\pi}[-2 A(M_{W})-5 q^{2}B_{0}(q^{2};0,M_{W})
-2 q^{2}B_{1}(q^{2};0,M_{W}) \\
&&-10 B_{22}(q^{2};0,M_{W})-\frac{2}{3}q^{2}+2 M_{W}^{2}], \\
\Sigma^{W}_{11}(q^{2})&=&\frac{\alpha}{4\pi}\frac{c_{w}^{2}}{s_{w}^{2}}
[-2 A(M_{W})-2 M_{Z}^{2}B_{0}(q^{2};M_{Z},M_{W})-5 q^{2}B_{0}(q^{2};M_{Z},M_{W}) \\
&&-2 q^{2}B_{1}(q^{2};M_{Z},M_{W})
-10 B_{22}(q^{2};M_{Z},M_{W})-\frac{2}{3}q^{2}+2 M_{W}^{2}+2 M_{Z}^{2}], \\
\Sigma^{W}_{12}(q^{2})&=&\frac{\alpha}{4\pi}\frac{1}{s_{w}^{2}}(3A(M_{W})-2M_{W}^{2}),\ 
\Sigma^{W}_{13}(q^{2})=\frac{\alpha}{4\pi}\frac{c_{w}^{2}}{s_{w}^{2}}
(3A(M_{Z})-2M_{Z}^{2}), \\
\Sigma^{W}_{14}(q^{2})&=&\frac{\alpha}{4\pi}\frac{c_{w}^{2}}{s_{w}^{2}}
B_{22}(q^{2};M_{Z},M_{W}),\ 
\Sigma^{W}_{15}(q^{2})=\frac{\alpha}{4\pi}\frac{c_{w}^{2}}{s_{w}^{2}}
B_{22}(q^{2};M_{Z},M_{W}), \\
\Sigma^{W}_{16}(q^{2})&=&\frac{\alpha}{4\pi}
B_{22}(q^{2};0,M_{W}),\ 
\Sigma^{W}_{17}(q^{2})=\frac{\alpha}{4\pi}
B_{22}(q^{2};0,M_{W}).
\end{eqnarray*}

\end{document}